\documentclass{article}

\usepackage{arxiv}
\usepackage{authblk}

\usepackage[utf8]{inputenc} 
\usepackage[T1]{fontenc}    
\usepackage{hyperref}       
\usepackage{url}            
\usepackage{booktabs}       
\usepackage{amsfonts}       
\usepackage{nicefrac}       
\usepackage{microtype}      
\usepackage{lipsum}
\usepackage{amsmath}
\usepackage{graphicx}
\graphicspath{ {./images/} }

\usepackage[style=authoryear-ibid,backend=biber]{biblatex}
\usepackage{float}
\title{Guitar Pickups I: Analysis of the Effect of Winding and Wire Gauge on Single Coil Electric Guitar Pickups}

\author[1]{C. Batchelor}
\author[1]{J. Gooding}
\author[2]{W. Marriott}
\author[2]{N. Chalashkanov}
\author[2]{N. Tucker}
\author[1,*]{R. Margetts}

\affil[1]{Department of Engineering, Nottingham Trent University, Nottingham, NG11 8NS, UK}
\affil[2]{School of Engineering, University of Lincoln, Lincoln, LN6 7TS, UK}
\affil[*]{Corresponding Author: rebecca.margetts@ntu.ac.uk}

\begin{document}
\maketitle
\begin{abstract}
Guitar Pickups have been in production for nearly 100 years, and the question of how exactly one pickup is tonally `superior' to another is still subject to a high level of debate. This paper is the first in a set demystifying the production of guitar pickups and introducing a level of scientific procedure to the conversation. Previous studies have analysed commercial off--the--shelf pickups, but these differ from each other in multiple ways. The novelty of this study is that dedicated experimental pickups were created, which vary only one parameter at a time in order to allow scientific study. The most fundamental qualities of a single-coil pickup are investigated: in this paper, number of turns and gauge of wire. A set of single--coil Stratocaster--style pickups were created, with the number of turns of wire varied across the commercially available range (5000 - 12000 turns), and this was done for two widely used wire gauges (42 and 44 AWG). A frequency response analyser was used to obtain impedance across a frequency range. It is shown that resonant frequency decreases exponentially with number of turns, while the magnitude of the resonant peak increases linearly with number of turns. The wire gauge used has a significant impact on both parameters, with the thicker wire giving higher resonant frequencies and higher magnitudes than the thinner wire for the same number of turns. These impact the sound associated with the pickup: the resonant frequency is linked to the perceived tone of the pickup, and the magnitude to the output amplitude and hence `gain.' Essentially, more turns will give a higher output pickup with a darker tone, while less turns will give a quieter pickup with a brighter tone -- consistent with what can be observed in commercial pickups. The choice of wire gauge alters the trade-off between the two, with the thicker 42 AWG wire giving louder outputs and brighter tones than the thinner 44 AWG wire. 
\end{abstract}


\section{Introduction}
\subsection{Background to Guitar Pickups}
The electric guitar has now been in popular use for nearly 100 years. Its popularity stems not only from its versatility but also from its lack of an abject requirement to develop an understanding of music theory, making it accessible to all. Unlike most classical instruments, a wide range of guitar designs are available and it is commonplace for players to modify them \parencite{Duchossoir}.

On an electric guitar, the sound produced is dominated by the function of the electric `pickups,' associated circuitry, and amplification (including the addition of any effects). Early electric guitars like the Rickenbacker "Frying pan" and Les Paul's "Log" barely resembled a guitar \parencite{Bacon, Owsinski}, and the contribution of the body, neck, and other hardware to any tonal properties -- while still hotly debated in the guitar community -- are widely recognised to be of far less importance than the electrical circuitry in the affect on amplified sound. A common modification to guitars is fitting after-market pickups, which can fetch over US\textdollar350 for a set: they might be replicas of the pickups found in a desirable vintage guitar, or have particular properties (like very high gain) to suit a style of playing.

Although there is some research and widely--understood guidance surrounding pickup design, conversations with pickup and custom guitar firms across the UK revealed that the precise effects and interplay between design variables are not well understood. Customers might request a very specific wire, for example, insisting that they can perceive a tonal difference that is not obvious to others.  

\begin{figure}
    \centering
    \includegraphics[scale=0.7,trim=2.5cm 0 0 0]{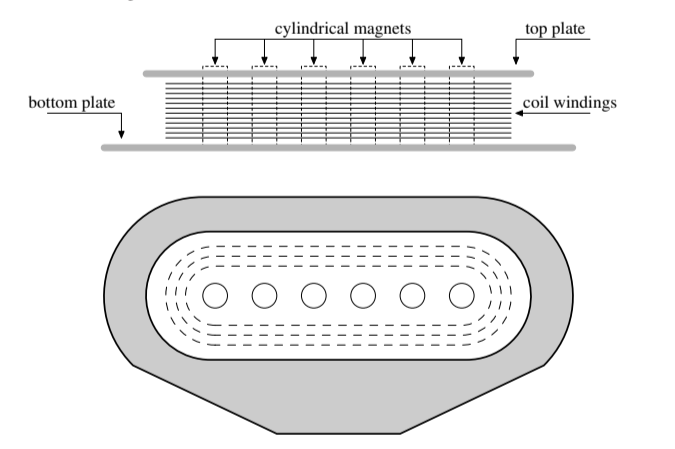}
    \caption{The Standard Construction of a Magnetic Pickup \parencite{Lahdevaara}}
    \label{fig:Pickup}
\end{figure}

A typical single-coil guitar pickup works as follows \parencite{Lahdevaara}. Copper wire is wound around a set of six `poles' which are magnets, or steel magnetised by an attached bar magnet, as seen in Fig. \ref{fig:Pickup}. This is held in place by face--plates and/or a bobbin, dipped in wax, and wrapped in tape to prevent damage to the fine wires of the coil. The pickup is positioned in the guitar's body directly under the six strings. For guitars with differing numbers of strings, such as bass guitars or seven--string guitars, the number of poles varies accordingly. The strings are always metal and they consequently become magnetised too. As the guitar is played, the strings vibrate and hence distort the magnetic field around them and the pole pieces. This induces an electromotive force to the coil and causes a current to flow. The coil is connected to the rest of the guitar's circuitry and ultimately the amplifier and speaker, to yield an audible sound. There is a wide array of pickup styles available: single and `humbucker' pickups, with varying number of turns of wire, wire gauge, type and coating, geometry, magnet composition and strength, layout of the wires (e.g. "scatter--winding"), and type of wax potting used to set the coil in place \parencite{oilcitypickups_2022_the}. The pickup can be thought of as a basic LCR circuit \parencite{Meinel_ch4}: the coils form an inductor, with distributed and stray capacitances between the wires and a resistance of the wire.

This is the first in a set of papers, the purpose of which is to establish the effects of the many design variables in a guitar pickup on tone, and to produce a usable guide for Luthiers and the after-market guitar hardware market. 

\subsection{Parameters of a Pickup}
It is well known among Luthiers and guitar enthusiasts that different pickups give very different sounds, and that this is related to certain characteristics like turns of wire, magnet choice, etc. The precise relationships and their importance are still unclear. 

The studies that have been conducted so far are limited: they crucially only test commercial off-the-shelf pickups with each other \parencite{Meinel_ch4, tian_2018_research, lemme_2013_is, jungmann_1994_theoretical, Novak, Guadagnin}. This is immensely problematic: there are a range of parameters affecting a pickup's output, and each commercial pickup varies several of these. So, comparing Gibson PAF to Gretsch Dynatron and Fender Stratocaster pickups will certainly yield different results, but the reader is left with no clue as to whether this is due to the number of turns of wire, significantly different geometries, or some other factor. To further cloud the issue, quality control on some pickups is famously lacking: original Fender pickups were reputedly wound for the length of time it takes to smoke a cigarette, using any materials available at the time. There can be significant variations in key properties between two pickups of even an identical brand and model. Existing studies also focus on electrical properties like impedance rather than relating tone to physical design choices like wire gauge, which means they are of limited use to a pick-up builder. 

Consequently, there is no specific guidance on the exact effects of each pickup parameter in isolation. These papers seek to establish this guidance, by creating sets of dedicated experimental pickups with one parameter varied across each set and all others controlled. These are then tested in a manner similar to the existing literature, but with the key difference that the effect of a single property on the measured outputs can be established. 

The focus of this paper will be the number of turns of wire on the pickup, as this will have an immediately obvious effect on the pickup's properties by altering the inductance. A well-known consequence of Faraday's Law is that electromotive force is increased if the number of turns in an electromagnet increases. \textcite{Meinel_ch4} identifies inductance as the most crucial property of a pickup, related in part to the number of turns (along with coil geometry and magnet/core geometry and properties). This study also incorporates a comparison of two common wire gauges, as there is not a sufficient variety of wire gauges in use to motivate a separate paper. 

In Part II, the geometry of the pickup will be explored. Wide, flat pickups like `pancake' and Jazzmaster pickups are associated with a distinctive tone. Then Part III will investigate the effects of magnet choice, which is a subject of much debate in guitar enthusiast circles. Further papers on the effects of wire material and potting are planned, concluding with a mathematical model and guidance aimed at the aftermarket pickup industry.

\subsection{Measurements}
 The existing literature concerning characteristics of a pickup often focuses on purely electrical characteristics such as DC Resistance, Impedance, Capacitance, and Inductance \parencite{tian_2018_research,Margetts2023}. \textcite{lemme_2013_is} argues that DC Resistance is "worthless," although \textcite{Meinel_ch4} notes its relationship to impedance and includes it in his model (which has since become widely-used and replicated). Both suggest that inductance is key and frequency response can be used to characterise a pickup. 
 
 Key literature uses frequency analysis and focuses on generating the impedance curves for given commercial pickups \parencite{Meinel_ch4,jungmann_1994_theoretical}. Impedance is a specific transfer function: the ratio of [output] voltage to [input] current \parencite{Nise}. It is established experimentally using a `swept sine' test where a sinusoidal input is applied and steadily increased in frequency while output is measured. A resonant peak(s) will be visible at the natural frequency (or frequencies). This work was originally conducted in 1987 and 1994 respectively, using a signal generator and millivoltmeter \parencite{Meinel_ch4,jungmann_1994_theoretical}. The resistor requires a high value e.g. \(200 k\Omega\): the authors believe the \(200 \Omega\) quoted in Meinel's book is a typographical error. There were high levels of variations evident in their results. Since then, commercial equipment and software for frequency analysis has become progressively better and more accessible. In addition to impedance, it is possible to measure pure gain (ratio of [output] voltage to an [input] voltage), or capacitance and reactance over the frequency range. The frequency range itself can be increased due to modern software allowing smaller test increments and larger sample sizes.

 \begin{figure}
    \centering
    \includegraphics[scale=0.6,trim=2.5cm 0 0 0]{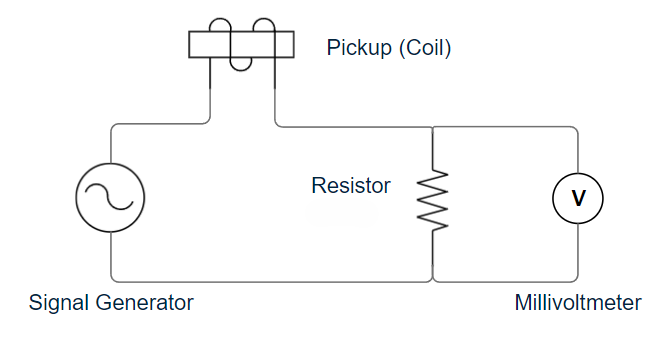}
    \caption{Apparatus for testing a Pickup \parencite{Meinel_ch4}}
    \label{fig:Test}
\end{figure}

 \subsection{Outputs linked to Perceived Tone}
 Generally speaking, the natural frequency of the pickup is thought to be correlated to tone. \textcite{Meinel_ch4} shows impedance curves for selected pickups, and Jungmann expands on this with a wider range of commercial pickups \parencite{jungmann_1994_theoretical}. It can be seen that pickups with a reputation for being "bright" or "glassy"  have a higher resonant frequency than those known for being "dark" or "singing." This has been replicated and expanded upon in luthiery resources e.g. \textcite{wacker_2019_mod} who attribute tonal descriptions and examples to resonant frequencies. For most of these pickups, there are clearly visible differences in geometry and well-publicised differences in the number of turns of wire. Other differences like the type of magnet, or wire gauge or coating, may not be as well-documented in the public domain. 
 
 In addition to the tone, the magnitude of the response is an important consideration. Lemme \parencite{lemme_2013_is} notes that pickups able to give a higher voltage output will drive an amplifier to distortion in different ways. Gain -- the relationship between input and output voltage -- may therefore be desirable in pickups for guitarists requiring a heavily distorted sound. He also states that a higher magnitude and more pronounced resonant peak "sound nicer." In classical instruments, the loudness can be perceived as quality \parencite{kitchen_2023_violin}, and this may be a factor in how a pickup's tone is perceived. 
 
\begin{figure}
    \centering
    \includegraphics[scale=0.6,trim=2.5cm 0 0 0]{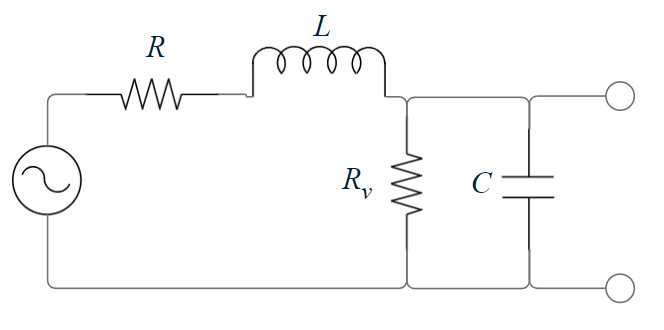}
    \caption{LCR Model approximating a Pickup \parencite{Meinel_ch4}}
    \label{fig:LCR}
\end{figure}

 However, the pickup is not a simple inductor. It has some resistance and capacitance too, and behaves as an LCR circuit shown in fig. \ref{fig:LCR}. It has inherent reactance / resistive losses, and distributed and stray capacitances: both also increase with number of turns. The impedance can be expressed in terms of resistance and reactances (capacitive and inductive). Neglecting the losses \(R_{v}\), this is a series LCR circuit expressed by eqn. \ref{impedance1}. At frequencies below resonance, the curve is dominated by capacitive reactance and behaves like a capacitor \parencite{jungmann_1994_theoretical}. At frequencies above resonance, the curve is dominated by inductive reactance and behaves as though it's an inductor. Increasing inductance creates a low-pass filter, which is why the bass and mid-range are increased relative to the treble.   
 
\begin{equation}
\label{impedance1}
    Z = \sqrt{R^2 + \left( X_{L} - X_{C} \right)^2}
\end{equation}
\begin{equation}
\label{impedance2}
    Z = \sqrt{R^2 + \left( \omega L - \frac{1}{\omega C} \right)^2}
\end{equation}
Where \(Z\) is impedance and \(X\) is reactance. The resonant peak seen on an impedance graph is the point at which capacitive and inductive reactances are equal, and \(Z = \sqrt{R^2}\). It occurs at a resonant frequency given by eqn. \eqref{resfreq}.
\begin{equation}
\label{resfreq}
    f_{res} = \frac{1}{2 \pi \sqrt{L C}}
\end{equation}

 The LCR model is supported by the observations of practitioners. \textcite{lemme_2013_is} notes that adjusting the volume potentiometer on a guitar can have a greater effect on treble than on bass, and shows frequency response (impedance) graphs for pickups with associated circuitry set to different volume settings. This work demonstrates that the \textit{shape} of the impedance curve is important in determining tone, not just magnitude or the resonant frequency. He suggests that increasing the number of turns of the coil, and hence increasing the inductance, should increase the bass and mid-range relative to the treble. This is consistent with the idea of high inductance creating a low-pass filter effect. 

\subsection{Wire Turns and Gauge}
The number of turns of wire around the pickup were identified as the most important property \parencite{Meinel_ch4}. The pickup is essentially an inductor -- a coil of wire -- and there are a number of factors to consider when winding an inductor \parencite{kuphaldt_2015_inductor}. These include the area of the coil and the permeability of the core, which will be investigated in future work relating to geometry and magnet choice. With regard to number of turns, Faraday's Law \eqref{faraday} shows that the electromotive force, or voltage, is increased when the number of turns on the coil is increased \parencite{french_2008_guitar} 
\begin{equation}
\label{faraday}
    e = -N \frac{d\phi}{dt}
\end{equation}
This increase in output means that impedance and gain are increased, and should therefore be visible on a frequency response graph as an increased magnitude. 

Increasing the number of turns increases inductance, which should be observable as a lower resonant frequency plus an increased magnitude (at least in the higher-frequency portion of the graph). However, it also increases resistance and capacitance: the graph shape will therefore alter.  

Wire gauge was identified as the second most effective factor in tone \parencite{oilcitypickups_2022_the}. There is little variation in the gauges of wire used (42-44 AWG) so the experiments were combined. This gives a relationship between number of turns and impedance for 42 AWG wire, and a second for 44 AWG wire. There is a 43 AWG wire (notably used on Telecaster neck pickups) and some vintage models will use 41 AWG: this was not available from the recommended supplier. 

The different wire gauges have different resistances: 42 AWG (0.063mm) has a resistance of 5.48 ohm/mtr at 20\(^\circ\)C, while 44 AWG (0.056mm) wire has a resistance of 6.94 ohm/mtr at 20\(^\circ\)C \parencite{Wires}. These are both figures for solderable enamelled copper wire, which is a standard choice. Other coatings are used: Formvar (on Vintage pickups and some replicas), Polyurethane and Poly Nylon, which will be investigated at a later date. 

The thinner wire (44 AWG) has a higher resistance and should therefore exhibit a lower impedance magnitude according to eqn. \eqref{impedance2}. Resistance also adds a damping effect, which may be visible as a wider peak. The wire has some capacitance, which increases with wire thickness.

\subsection{Hypotheses and Aim}
It is hypothesised that varying the number of turns around the coil will have a two-fold effect on the tone. 
\begin{enumerate}
    \item Increasing number of turns will decrease the frequency of the resonant peak seen on an impedance plot. This is indicated by the comparisons of commercial pickups conducted by \parencite{Meinel_ch4,jungmann_1994_theoretical} and Lemme's observations \parencite{lemme_2013_is}. 
    \item Increasing number of turns will increase the voltage induced, and hence the magnitude of the output. This may not be a simple change in magnitude but a change in the shape of the curve.
\end{enumerate}

It is also hypothesised that the different wire gauges will give different frequency response curves, as the thicker wire should have a lower resistance and higher capacitance than the thinner one.  

The aim of this paper is to obtain quantified relationships between physical variables (no. turns, gauge of wire) and indicators of tone (amplitude of impedance and resonant frequency). This can be formulated into guidelines for the aftermarket pickup and custom guitar industries.

\section{Materials and Methods}
A set of experimental pickups were wound to a known set of specifications. This allowed one parameter to be varied while all others were controlled. A classic Fender Stratocaster single coil pickup was selected for simplicity and repeatability, and constructed in line with guidance from Oil City \parencite{oilcitypickups_2022_the}.

Two sets of pickups were created, one with the 42 AWG wire and one with 44 AWG wire. 42 and 44 AWG solderable enamelled copper wires were selected, as these are commonly used. They were purchased from a recommended supplier \parencite{Wires} due to known variabilities in manufacturing quality from others. There is a 43 AWG wire which was neglected here, as are other wire coatings. Each set comprised fifteen pickups, with numbers of turns varying from 5000 turns (the lowest seen commercially: consistent with a Gibson PAF) to 12000 turns (in excess of most commercial single-coil pickups) in 500-turn increments. 

All other parameters in the pickup were controlled. The pickups were all fitted with AlNiCo V rod magnets (common on this type of pickup) of a height distribution standard on Stratocaster pickups, and of the same orientation. The flatwork was lasercut from vulcanised fibreboard. Other parts and consumables like eyelets, solder, flux, and glue were controlled and in line with Oil City's recommendations. The pickups were not potted (dipped in warm wax, as is the norm) to allow a future comparison between the properties of potted and un--potted pickups. This resulted in them being particularly fragile, and they were stored in foam-padded trays.

The pickups were wound on a commercially-available programmable pickup winder: the CNC Guitar Pickup Mini Coil Winder from CNC Design Limited \parencite{CNC}. This allowed the pickups to be wound in an approximately comparable manner. Note that it is widely thought that hand-winding and "scatterwinding" pickups gives a more pleasing tone, but the authors were aiming for scientific comparison.

\begin{figure}
    \centering
    \includegraphics[scale=0.1]{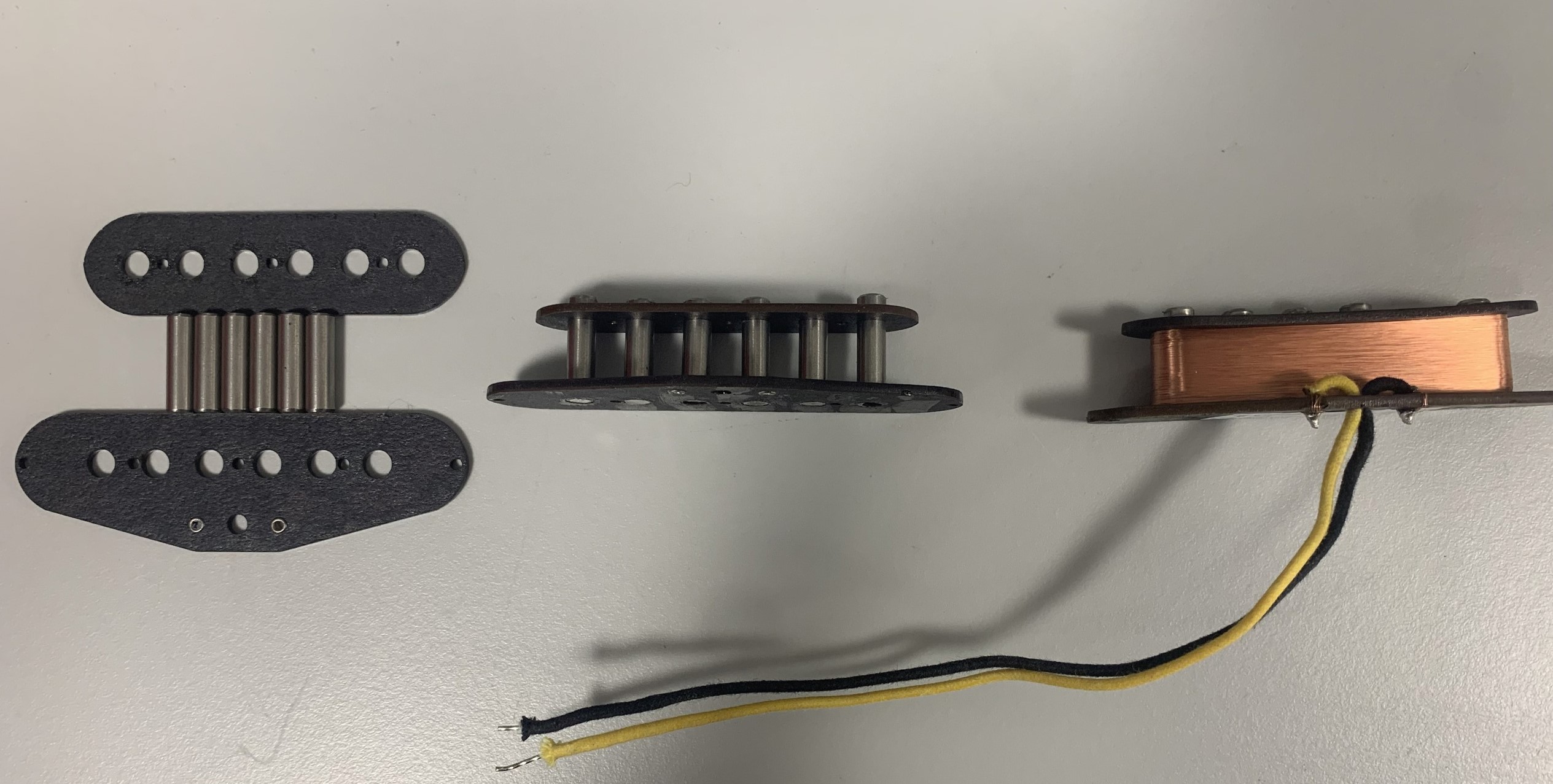}
    \caption{Three Stages of a Pickup's Production}
    \label{fig:Production}
\end{figure}

When the pickups had been wound, they were sent to the University of Lincoln's Engineering Department for testing. The Materials Laboratory is furnished with a VersaSTAT 4 \parencite{Versastat} with frequency response analyzer. This was used to provide impedance analysis over the frequency range 0-25 kHz (consistent with that recommended by \textcite{Meinel_ch4}. This gives plots of impedance against frequency consistent with the studies of commercial off-the-shelf pickups by \textcite{Meinel_ch4,jungmann_1994_theoretical} and the analysis promoted by \textcite{lemme_2013_is}. Each pickup was tested in isolation, without any of the surrounding circuitry found on an electric guitar. 

\section{Results}
\begin{figure}[ht]
    \centering
    \includegraphics[scale=0.3,trim=2.5cm 0 0 0]{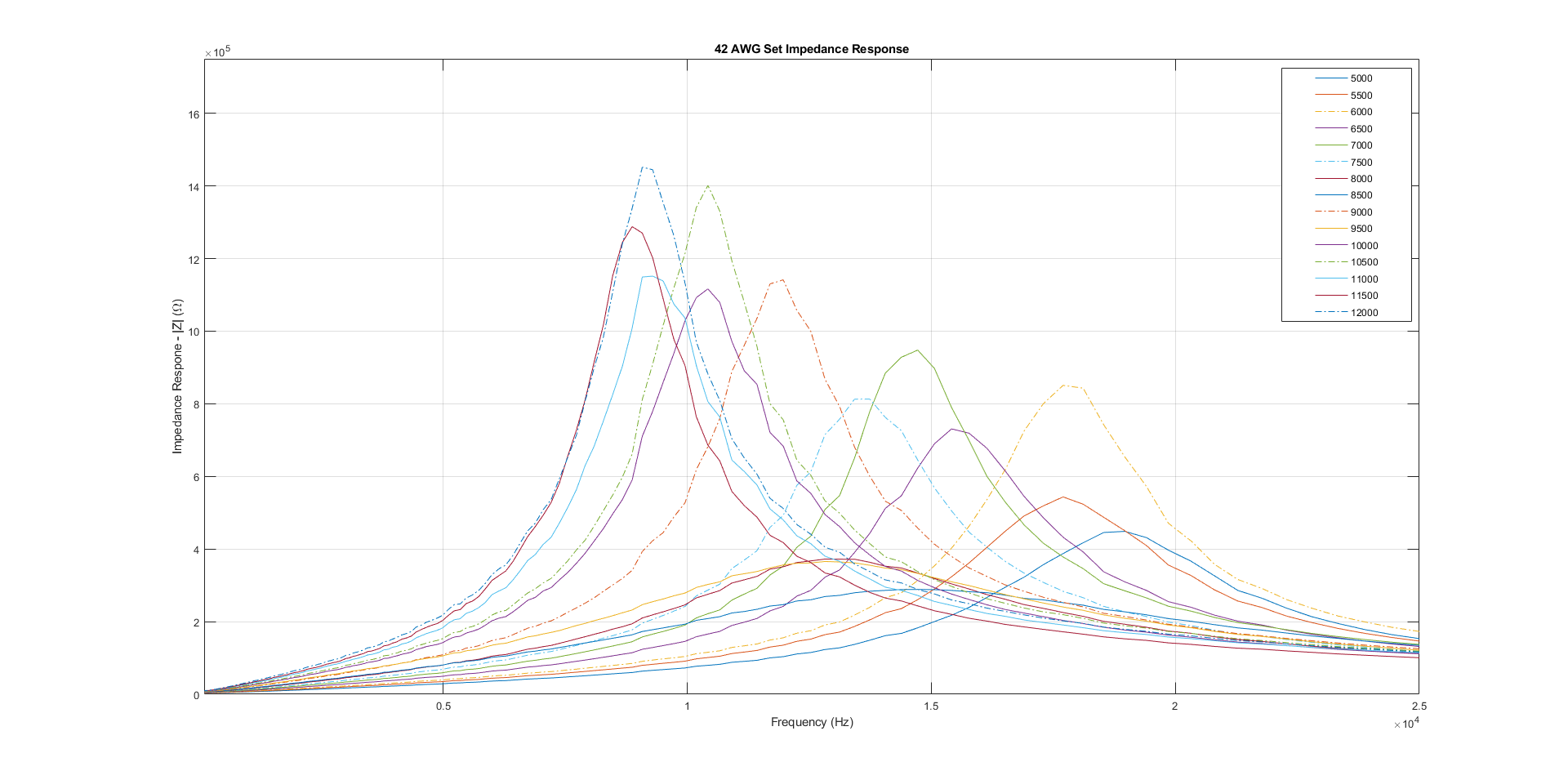}
    \caption{Impedance Response for varying number of turns - 42AWG}
    \label{fig:IRvNoT42}
\end{figure}

\begin{figure}[ht]
    \centering
    \includegraphics[scale=0.3,trim=2.5cm 0 0 0]{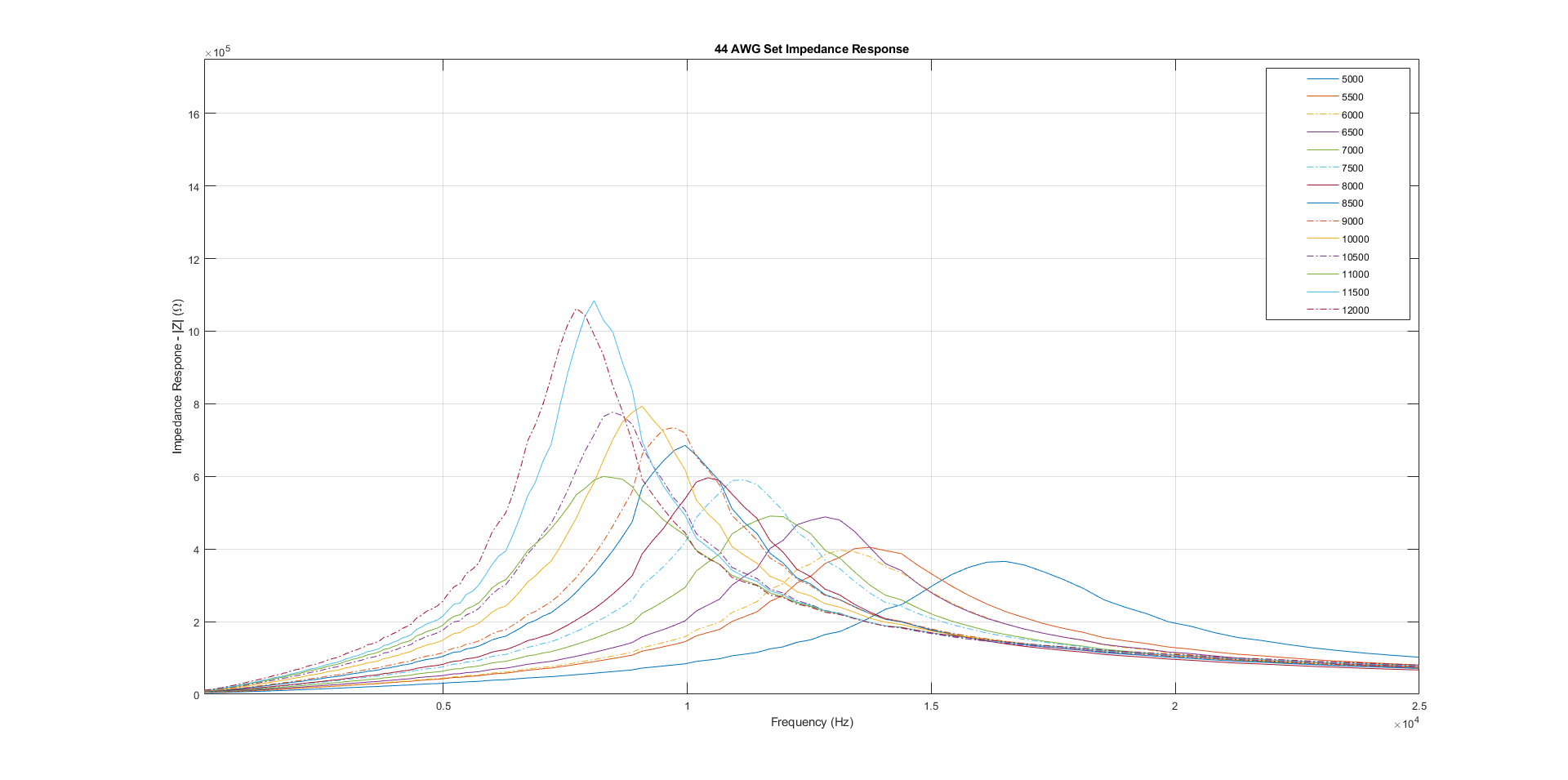}
    \caption{Impedance Response for varying number of turns - 44AWG}
    \label{fig:IRvNoT44}
\end{figure}

  Figs. \ref{fig:IRvNoT42} and \ref{fig:IRvNoT44} show that the amplitude of impedance generally increases with number of turns (with some variation), whereas the resonant frequency decreases exponentially. It can be assumed that as the number of turns decreases the resonant frequency approaches a minimum amplitude.  

\begin{figure}[ht]
    \centering
    \includegraphics[scale=0.3,trim=2.5cm 0 0 0]{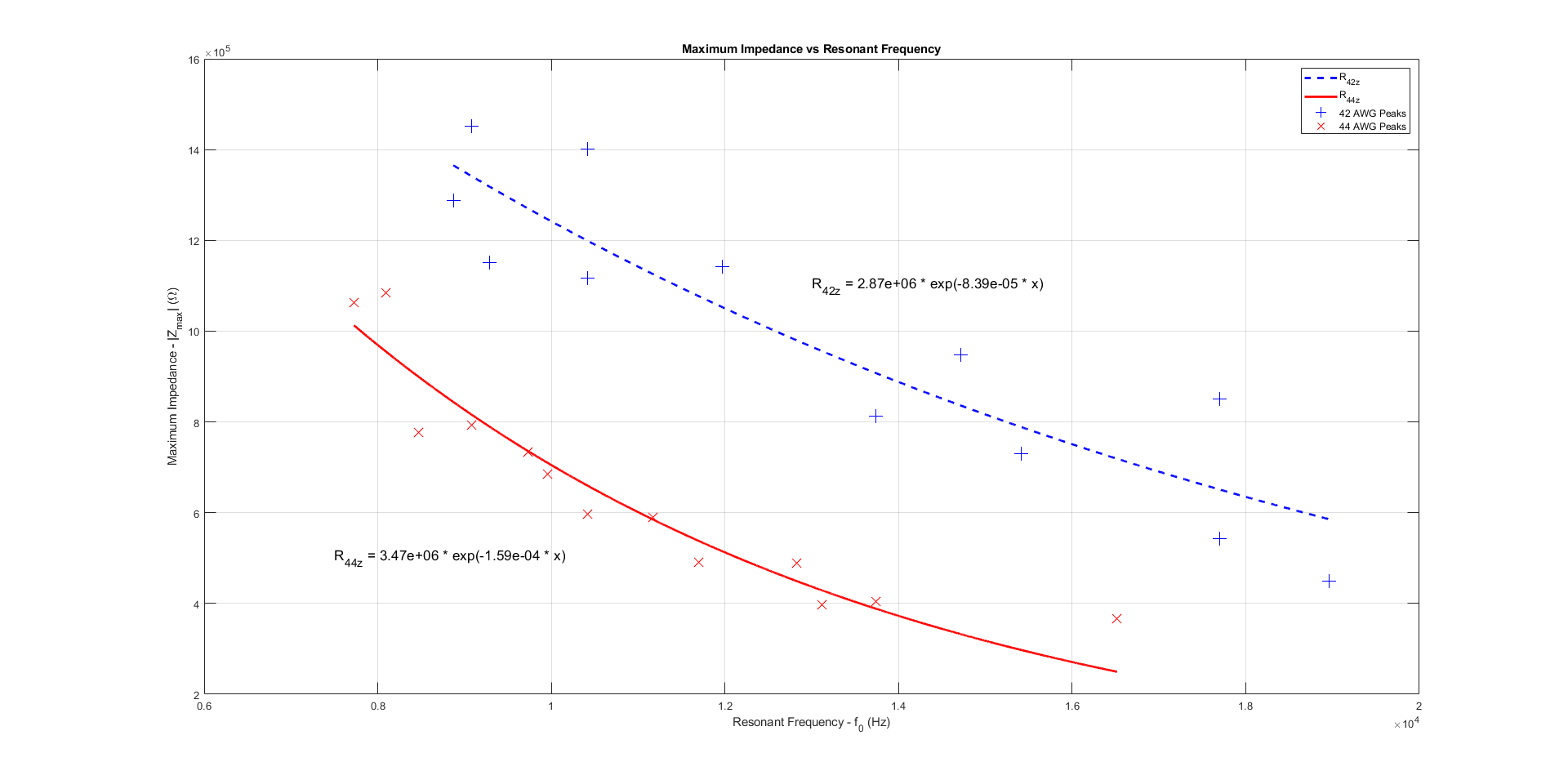}
    \caption{Maximum Impedance vs Resonant Frequency, including regression lines}
    \label{fig:42v44IRvsRes}
\end{figure}

Fig. \ref{fig:42v44IRvsRes} shows the relationship between maximum impedance recorded (i.e. magnitude of peak) and the frequency at which the resonant peak occurs. Number of turns fundamentally affects inductance, both increasing the output (and hence magnitude) and shifting the resonant frequency downwards. Note that the thicker 42 AWG wire has both significantly higher maxima and slightly higher resonant frequencies than the thinner 44 AWG wire.

\begin{figure}
    \centering
    \includegraphics[scale=0.3,trim=2.5cm 0 0 0]{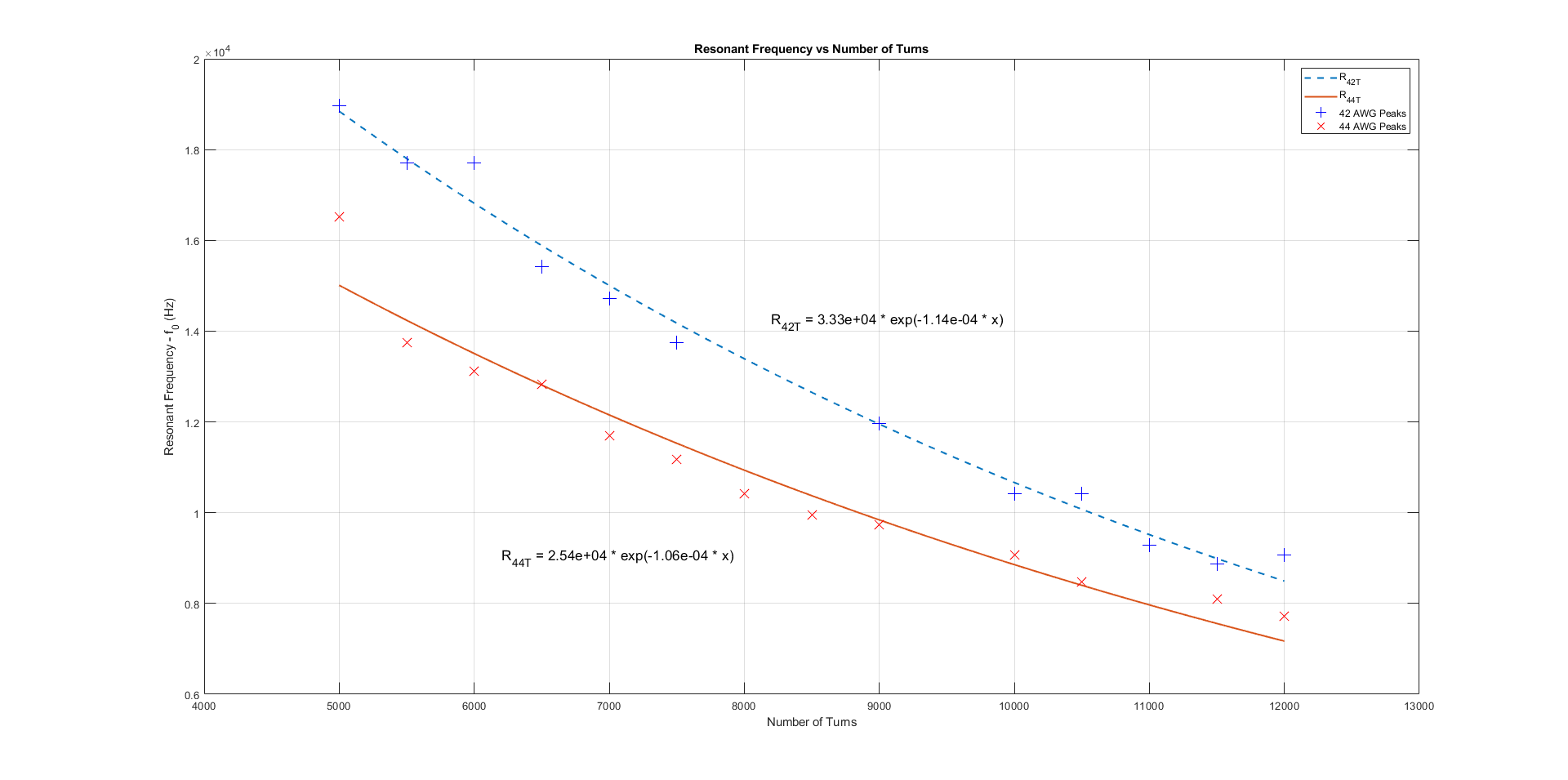}
    \caption{Resonant Frequency vs Number of Turns, including regression lines}
    \label{fig:42v44ResvNot}
\end{figure}

Fig. \ref{fig:42v44ResvNot} shows the relationship between number of turns and resonant frequency. This was plotted so as to create a guideline for practitioners who build guitar pickups. The thicker 42 AWG wire can be seen to give higher resonant frequencies than the thinner 44 AWG wire for the same numbers of turns. In both cases, resonant frequency decreases exponentially as number of turns increases.

\begin{figure}
    \centering
    \includegraphics[scale=0.3,trim=2.5cm 0 0 0]{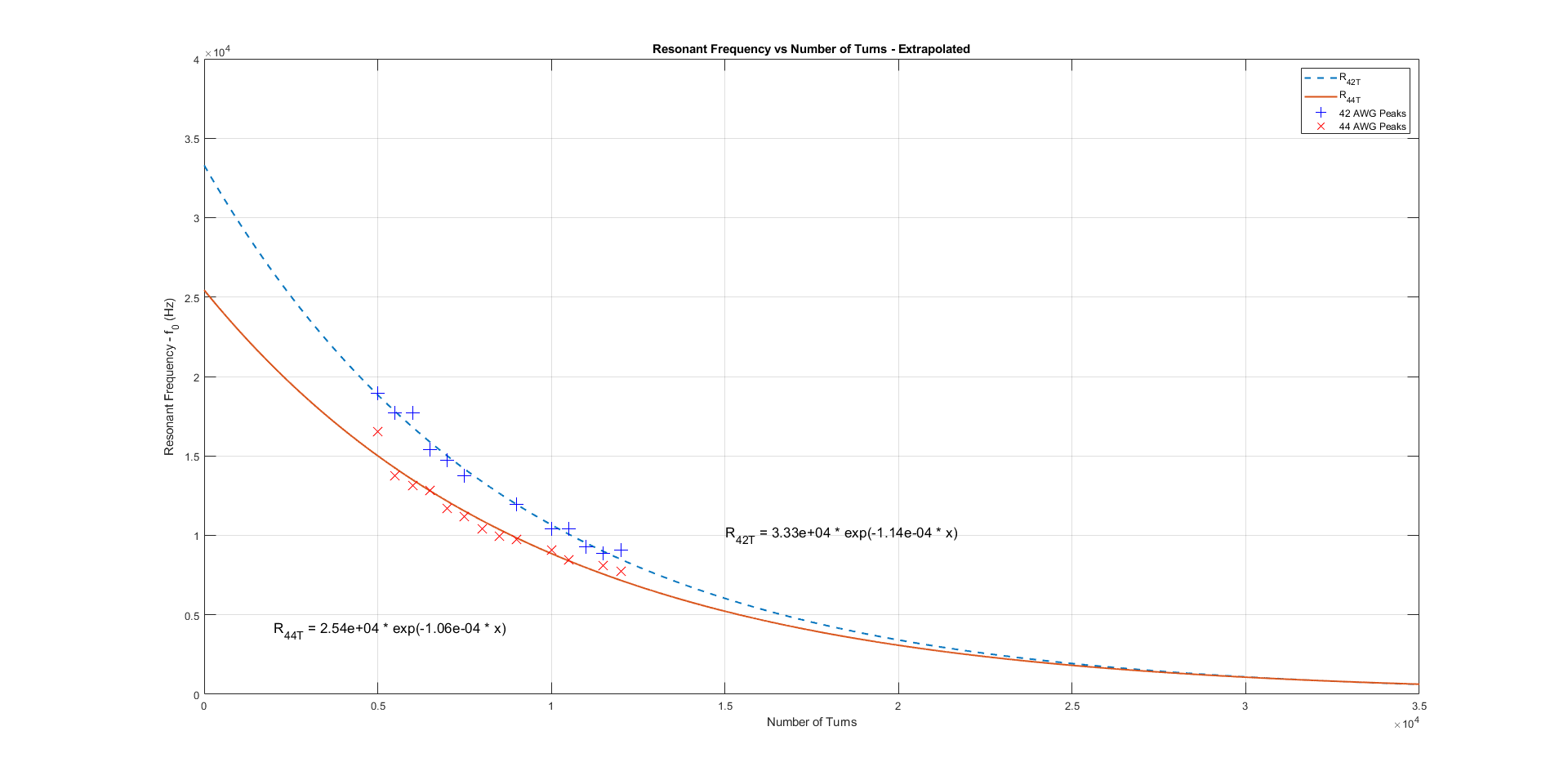}
    \caption{42 and 44 AWG Resonant Frequency vs Number of Turns, including extrapolated regressions}
    \label{fig:42and44ExtrResvNot}
\end{figure}

Fig. \ref{fig:42and44ExtrResvNot} extrapolates this relationship to predict resonant frequencies from 0 to 35000 turns.

\begin{figure}
    \centering
    \includegraphics[scale=0.3,trim=2.5cm 0 0 0]{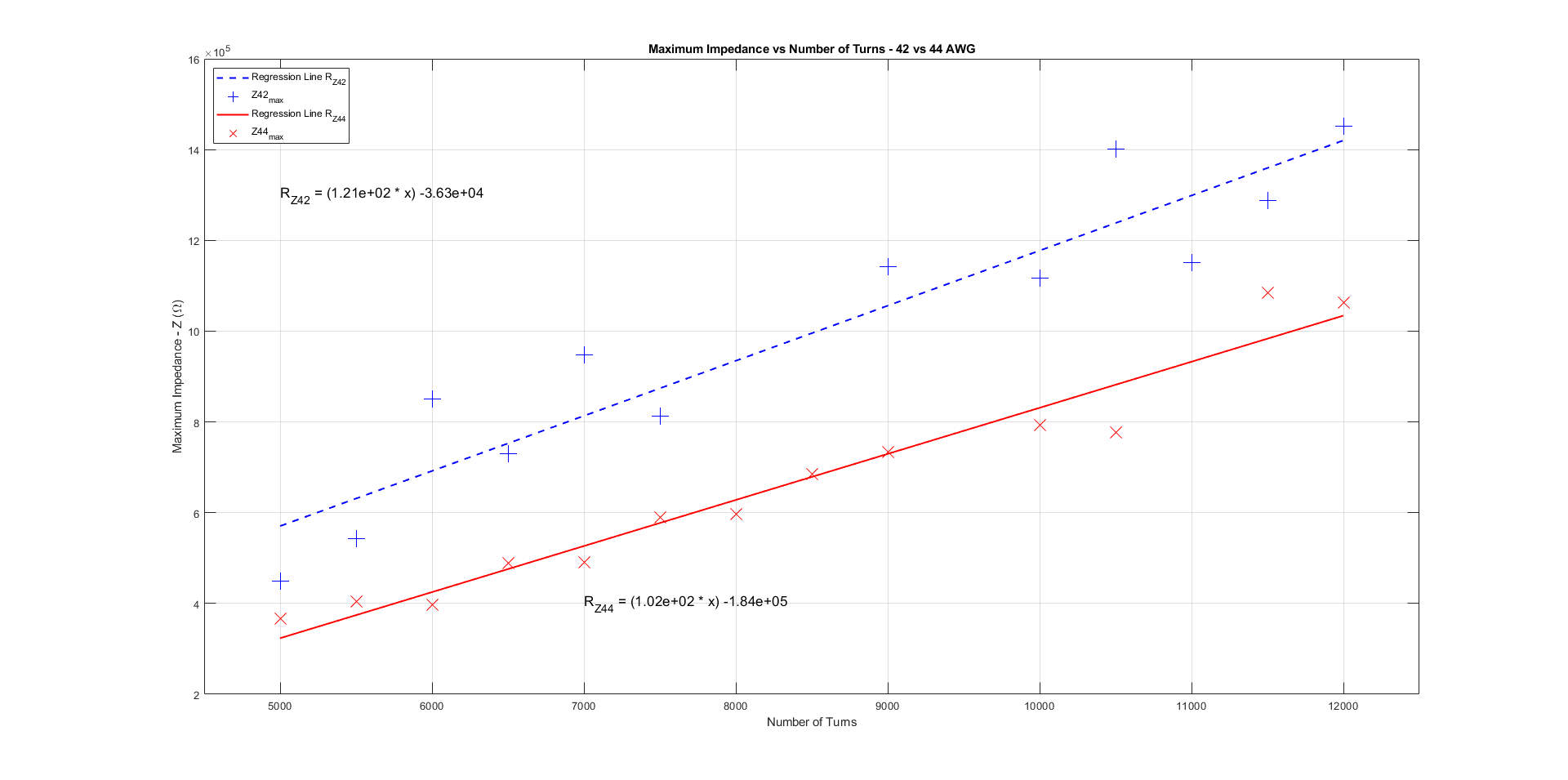}
    \caption{Maximum Impedance vs Number of Turns}
    \label{fig:42v44MaxImpvNoT}
\end{figure}

Figure \ref{fig:42v44MaxImpvNoT} shows the positive linear relationships between maximum impedance and number of turns of wire. Increasing number of turns does increase the magnitude of the impedance and, by inference, the output. Amplitudes are significantly higher for the thicker 42 AWG wire than for the thinner 44 AWG one.

\section{Discussion}
From these results, it is confirmed that increasing number of turns will decrease the frequency of the resonant peak seen on an impedance plot.  Figs. \ref{fig:IRvNoT42} and \ref{fig:IRvNoT44} show the full impedance curves for each wire gauge respectively, and an exponential relationship is visible. Fig. \ref{fig:42v44ResvNot} gives the precise relationships for each wire gauge, stated in eqn.s \eqref{Res42AWG} and \eqref{Res44AWG}. In both cases, resonant frequency decreases exponentially as number of turns increases. The relationship can be extrapolated, as in Fig. \ref{fig:42and44ExtrResvNot}, although there are practical limitations in that zero turns will clearly not yield an output, and large numbers of turns will be difficult to fit on a conventional pickup.

\begin{equation}
\label{Res42AWG}
    \text{42AWG: }f_{0} = 3.33 \times 10^{4} e^{-1.14\times10^{-4}N}
\end{equation}
\begin{equation}
\label{Res44AWG}
    \text{44AWG: }f_{0} = 2.54 \times 10^{4} e^{-1.06\times10^{-4}N}
\end{equation}

In addition to the frequency of the resonant peak, the shape of the response varied with number of turns. A wide response and lower amplitude are seen with fewer turns, and a narrower peak and higher amplitude are observed with more turns. As natural frequency decreases, a gradual increase in overall impedance and in gradient on the upper side of the peak is observed, consistent with increasing inductance.

It is also evident from figs. \ref{fig:IRvNoT42} and \ref{fig:IRvNoT44} that increasing number of turns will increase the magnitude of the impedance. Since the impedance is the ratio of output to input, and the input amplitude remains constant, this is due to a higher output amplitude. Increasing the number of turns therefore yields a higher output. The maximum value (i.e. magnitude of the resonant peak) varies linearly as shown in \ref{fig:42v44MaxImpvNoT} and according to eqn.s \eqref{Max42AWG} and \eqref{Max44AWG}.

\begin{equation}
\label{Max42AWG}
    \text{42AWG: } |Z_{f_0}| = 121 N - 3.63\times10^{4}
\end{equation}
\begin{equation}
\label{Max44AWG}
    \text{44AWG: } |Z_{f_0}| = 102 N - 1.84\times10^{5}
\end{equation}

These equations suggest that impedance will become negative at a low number of turns, which is clearly not possible. Further testing could identify the precise relationship, but for practical purposes pickups with under around 2000 turns are unlikely to give a usable response. 

Fig. \ref{fig:42v44IRvsRes} shows the relationship between maximum impedance recorded (i.e. magnitude of peak) and the frequency at which the resonant peak occurs. Both of these these things vary as a result of inductance, which is a consequence of number of turns -- hence the strong correlation.  

The two different wire gauges do give different frequency response curves. The thicker (42 AWG) wire gives higher maximum impedances (and consequently output amplitudes), likely due to having less resistance. The thicker 42 AWG wire can be seen to give higher resonant frequencies for the same number of turns. Resistance does not affect the resonant frequency as shown in eqn. \eqref{resfreq}, so this is likely due to increased capacitance.

From these results, it is clear to see that the initial hypothesis of increased amplitude with number of turns is accurate, as is that of decreased natural frequency with number of turns. This suggests the assertion by \textcite{lemme_2013_is} that a pickup with more turns will have a higher bass- and mid-range response is accurate.

Although the trends are as expected, they merit further comparison with the known literature as is it not clear how these figures relate to sound or tone. 

A vintage pre-1965 Fender Stratocaster pickup (comparable in style to the ones built here) uses 42 AWG wire and approximately 8350 turns \parencite{Vintage}. According to Figure \ref{fig:42v44ResvNot}, this pickup would have a natural frequency of \(3.33 \times 10^{4} e^{-1.14\times10^{-4}8350} \approx 12.9 kHz\). This is roughly comparable to the SSL--1 vintage--correct Statocaster pickups analysed by \textcite{jungmann_1994_theoretical}, which had resonant frequencies around \(11 kHz\). The discrepancy may be due to wire coating and winding pattern: the SSL-1's use Formvar--coated wire and are scatterwound, which will both likely affect the capacitance \parencite{SeymourD}). 

The trend noted here is consistent with the literature. Fender Stratocasters are known for their 'bright' tone with an emphasized top-end, and it's been noted that when Fender started using fewer turns under CBS-ownership from 1965, the pickups sounded brighter and weaker than those from the early 60s  \parencite{Guitar}. They also switched to enamelled wire and machine winding at that time \parencite{Vintage}. \textcite{jungmann_1994_theoretical} records lower natural frequencies for Humbucker (\(7.3 - 8.7 kHz\)) and P90 Soapbar (\(5.2 kHz\)) pickups, which are associated with `warm' and `dark' tones, while the De Armand / Dynasonic 1951 Gretsch pickups, known for clear tones and `twang,' had much higher resonant frequencies (\(13 - 24 kHz\)).

Note that the natural frequencies obtained here and by \textcite{jungmann_1994_theoretical} are significantly different to those obtained by others \parencite{Meinel_ch4,lemme_2013_is,wacker_2019_mod}: they report Stratocaster pickups as having natural frequencies around \(5 kHz\). This may be due to other variations in the pickup build, and/or differences in test equipment / setup. For example, \textcite{jungmann_1994_theoretical} notes significant influences of connecting cable capacitance and input resistance, and obtains different frequencies with different equipment. However, all sources do note similar relationships with `warmer' off-the-shelf pickups having lower natural frequencies than `brighter' ones. Therefore, the discrepancy in frequency values obtained does not preclude drawing conclusions about the general relationship between wire gauge, turns, and tone (based on change in measured natural frequency). 
    
Assuming the widely held notion that natural frequency correlates to tone is correct (to be tested in future work), the tonal properties of the pickup can be deduced from these relationships. It can be surmised: 

\begin{enumerate}
    \item Thicker (42 AWG) wire yields generally higher natural frequencies and hence `brighter' tones than thinner (44 AWG) wire, which yields lower natural frequencies and hence `warmer' tones.
    \item Pickups with higher numbers of turns yield lower natural frequencies and hence `warmer, darker' tones than those with fewer turns. 
\end{enumerate}

Linear relationships between magnitude (value of the impedance maxima occuring at the resonant frequency) and number of turns have been discovered and quantified for 42 and 44 AWG enamelled copper wire. 
\begin{itemize}
    \item Increased number of turns (i.e. more material) provides higher magnitude impedance. 
    \item Thicker wire (i.e. more material laid per turn) leads to higher impedance magnitude and a steeper increase with no. turns.
\end{itemize}

In acoustic instruments, loudness is associated with quality \parencite{kitchen_2023_violin} but this does not necessarily translate to electrical instruments. This is a potential new avenue of future research.

There is some variation in the results which can be attributed to some or all of the following observations: 
\begin{itemize}
    \item Some bobbins had shifted slightly during the gluing/setting process so the standard height that was set within the software did not cover the entire exposed section of pole pieces plate-to-plate. If this occurred then the number of completed winds was noted down, the winder re-zeroed, the number of winds in the system reduced by the number of completed winds, and then the winder started again. This led to minor discrepancies in the specific order of winding of the pickups.
    \item The two batches of magnets that were purchased were acquired from different companies, the initial batch being unmagnetised and slightly thicker magnets, the latter being pre-magnetised and slightly thinner. This may lead to variations in the magnetic strength of the pickup. This should not have had any significant effect on the coil properties investigated here. 
    \item Due to the machine tolerances of the laser cutter, not every magnet hole was perfectly evenly cut, which led to a lot of the magnets settling at slightly skewed angles. Again, this should not have had any significant effect on the coil properties investigated here but will be considered in future work on the effects of magnet and magnetic field.
    \item The alignment of the CNC winder can take several attempts to be completed properly, leading to minor variation between the indicated number of winds and the true number of winds, though this was never more than 5 attempts (10 winds) and so was considered insignificant. 
\end{itemize}

\section{Conclusions}
This study confirmed that varying the number of turns of wire on the pickup will affect parameters associated with tone. 

Increasing number of turns decreases the frequency of the resonant peak seen on an impedance plot. Resonant frequency is related to number of turn by eqn.s \eqref{Res42AWG} and \eqref{Res44AWG}. 

The nature of this relationship is consistent with the literature, although the frequencies obtained are different to some sources (likely due to differences in test equipment). The frequency of this resonant peak has been linked to tone in the literature, so it follows that this relationship describes tonal variation; fewer turns give a higher natural frequency and a `brighter' tone, whereas more turns give a lower natural frequency and a `darker' tone. 

Increasing number of turns increases the voltage induced. This is evident from the increase in impedance magnitude, which can only be due to an increased output. There is a positive linear relationship between number of turns and maximum impedance value shown by eqn.s \eqref{Max42AWG} and \eqref{Max44AWG}. The maximum impedances are recorded at differing frequencies.
    
This is not a simple change in magnitude but a change in the shape of the curve. A wide response and lower amplitude are seen with fewer turns, and a narrower peak and higher amplitude are observed with more turns. As natural frequency decreases, a gradual increase in overall impedance and in gradient on the upper side of the peak is observed. This is because resistance and capacitance of the wire factor into the shape of the curve. 

Different wire guages give significantly different frequency response curves. The 42 AWG wire is thicker than the 44 AWG, giving it a lower resistance and hence higher maximum impedances (and consequently output amplitudes). The thicker 42 AWG wire also gives a higher capacitance, and therefore gives higher resonant frequencies than the 44 AWG wire for the same number of turns. 

The aim of this paper has been achieved. Quantified relationships between physical variables (no. turns, gauge of wire) and indicators of tone (amplitude of impedance and resonant frequency) are established and can be formulated into guidelines.

\subsection{Further Work}
Part II of this set of papers will look at pickup geometry, as this is likely to affect impedance and hence tone. 

Part III will look at the effects of the magnets used: something Oil City \parencite{oilcitypickups_2022_the} consider the third most important property. This is perhaps a controversial study: some guitar enthusiasts highly value specific magnets, while others deny any impact on tone. \parencite{lemme_2013_is} states ``There is no Alnico 2 sound, nor an Alnico 5 sound..." 

Another intriguing aspect of the subject is the arrangement of layers during winding, so it is proposed that a further research project makes use of the Standard Script Engine provided in the winding software. This provides far greater control over the winding process and would allow for the introduction of predetermined patterns into the winding process, as opposed to the continually winding, even-layered, back-and-forth protocol displayed by the standard Automated Winding process. 

Further work is also proposed analysing audio signals from these pickups, including through DSP methods and with subjective audio testing. The pickups can then be tested in their normal mode of operation and analysed appropriately. This will necessitate consideration of suitable repeatable inputs, such as an ebow or measuring strum force. This would also finally establish a link between resonant frequency / output magnitude and perceived tone. 

\section*{Acknowledgements}
This work was funded through the `Engineering Research Opportunities Scheme' (EROS) at Nottingham Trent University, and the `Undergraduate Research Opportunities Scheme' (UROS) at the University of Lincoln. These competitive schemes allow promising undergraduate students to conduct research with members of staff over the summer. Additional `pump priming' equipment and training was funded by the Department of Engineering at Nottingham Trent University.

The specific contributions of the authors are as follows. 
\begin{itemize}
    \item Charles Batchelor secured EROS projects in 2022, 2023 and 2024 during which he was trained to wind pickups by Oil City Pickups in London, created a set of experimental pickups, analysed results to establish trends, and wrote the first draft of the paper. 
    \item Jack Gooding secured an EROS project in 2023 where he worked with equal contribution to Charles that year in creating a set of experimental pickups and analysing the results. 
    \item William Marriott secured a UROS project in 2023 and was responsible for testing the pickups using the frequency response analyzer. 
    \item Nikola Chalashkanov proposed the use of the frequency analyser, offered technical expertise in the field of electromagnetism, and joint-supervised William. 
    \item Nick Tucker suggested the collaboration, hosted all authors at Lincoln's Materials and Metrology labs, offered technical expertise on the wire material used, and joint-supervised William. 
    \item Rebecca Margetts initially proposed the project in 2022 as part of a wider body of work on stringed instrument design in the newly--formed Music Engineering Lab. She supervised Charles and Jack, was trained to wind pickups by Oil City Pickups in London, offered expertise in the field of modal analysis and signal processing, conducted the initial literature review, and worked on the final draft.
\end{itemize}
All authors were given the final draft for review and approval.  

The authors would like to acknowledge to contributions of a range of colleagues. Tim Bailey and Ash Scott-Lockyer of Oil City Pickups generously shared their expertise and observations, and confirmed the impact this work will have on boutique manufacturers such as Oil City. Dave Turner (currently of Make Noize Studios in Sheffield) contributed further literature and his experiences as a studio engineer and guitar enthusiast. Members of the NTU Technical team supported work in the laboratories and contributed their experiences of luthiery and guitar customisation. 

There is no conflict of interest to declare. No Artificial Intelligence was used in this work or the preparation of this paper.

This article will be submitted for review to the AES Journal, October 2024.

\printbibliography

\end{document}